\documentclass[11pt,paper,amsmath,amssymb,bbl]{JHEP3}
\newcommand\fverb{\setbox\fverbbox=\hbox\bgroup\verb}
\newcommand\fverbdo{\egroup\medskip\noindent%
			\fbox{\unhbox\fverbbox}\ }
\newcommand\fverbit{\egroup\item[\fbox{\unhbox\fverbbox}]}
\newbox\fverbbox

\usepackage{epsfig,multicol,bbm,graphics,graphicx}
\title{In search of the true vacuum: natural ordering, $\gamma$ condensate and the last renormalization}

\author{B. Pomori\v sac,\\
The "Vin\v ca" Institute of Nuclear Sciences, Laboratory for
theoretical and condensed matter physics --020, P.O. Box 522, 11001
Belgrade, Serbia \\ E-mail: \email{bvp@vin.bg.ac.yu}}

\abstract{With the idea of canceling the leading divergence in
vacuum  energy of $\varphi^4$ field theory a parameter is introduced
that interpolates between free Hamiltonian with or without normal
ordering. This leads to a condensate ground state having an
arbitrary number of particle-particle pairs. In addition to the
usual states, the condensate supports the states of negative energy
and negative norm. An explicit expression for the condensate state
is derived and perturbation theory with this state investigated. The
propagator is modified off the mass shell while unchanged on the
mass shell. Lowest order correction to the vacuum energy is
calculated and conditions for cancelation of the leading divergence
investigated. One possible solution is that all radiative
corrections in this formulation vanish. The other possible solution
implies a phase transition above the coupling of
$\frac{(2\pi)^2}{3}$ and the condensate non-analytical in the coupling constant. Possible implications are discussed.}

\keywords{Renormalization, regularization and renormalons, Nonperturbative Effects}
\dedicated{Dedicated to\ldots\\if you want.}
\preprint{}
\begin{document}


\section{Introduction and Motivation}
\par Let us start with a simple pedagogical example. Consider the motion
of a particle in a potential well shown\footnote{This potential for
a quantum mechanical particle is zero space and one time dimensional
analog of the Higgs potential.} in a figure 1.  Assume that the
potential is not precisely known to a physicist and he/she uses
parabolic, or quadratic approximation.
\FIGURE[h]{\epsfig{file=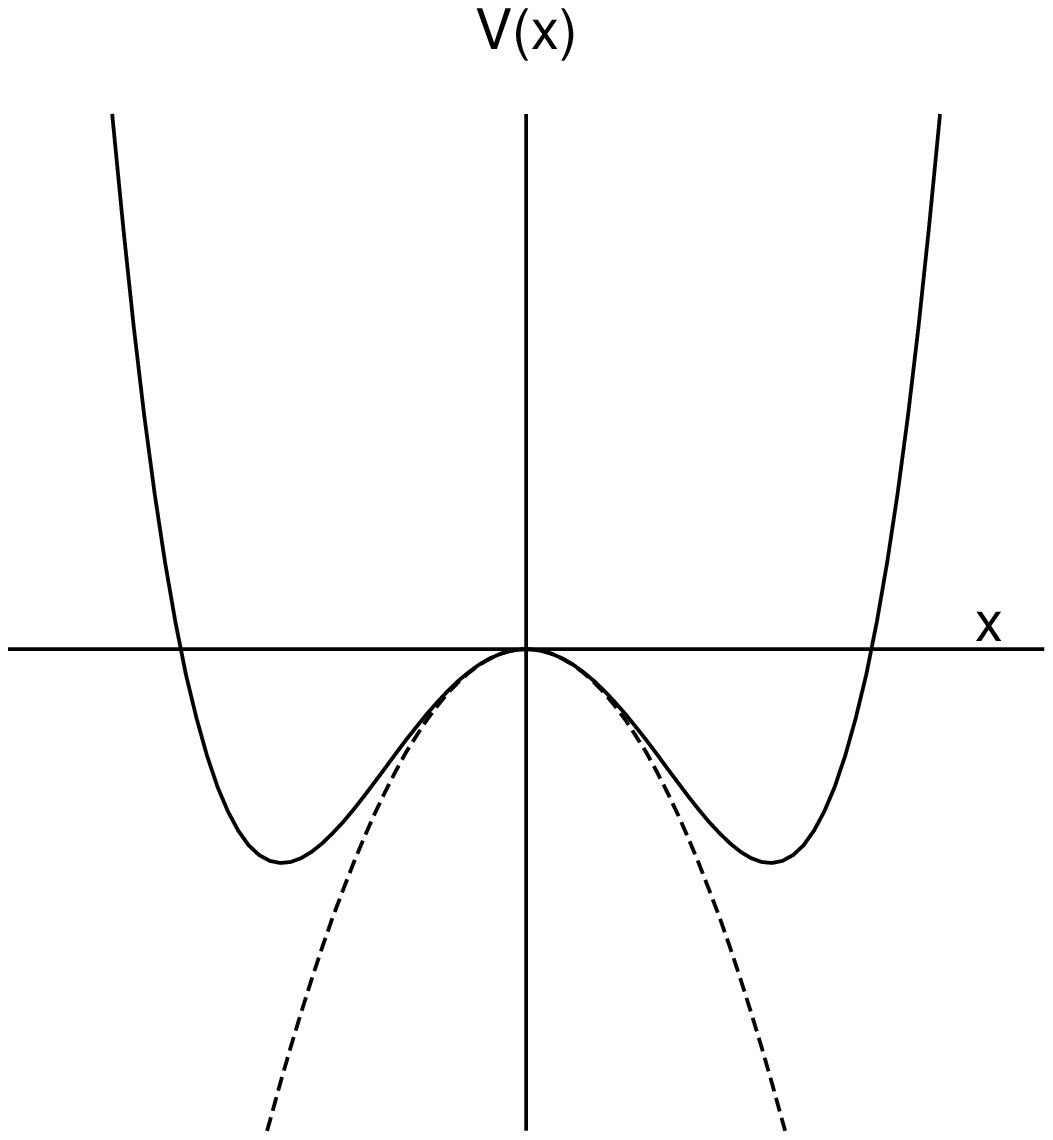,height=5cm}
        \caption{Potential for a particle motion full line, quadratic approximation to the potential dashed line}
	\label{fig_1}}

The physicist expands around the origin believing it to be a
minimum. It is not and they are ample signs of it. Among others, the
correction to the energy calculated in the said approximation is
infinite. Indeed, the divergent correction to the energy is standard
sign of expanding around the wrong "minimum". That is so everywhere
in physics, everywhere that is, except in Quantum Field Theory
(abbreviated QFT). There one accepts the infinite energy and
consequently speaks of the cosmological constant
problem \cite{zel,wei,car,pad}. The problem is, that an infinite, or
very large, vacuum energy is incompatible with the general
relativity where energy is the source of gravitational effects. The
gravitational effects of a large vacuum energy have not been seen
and something is wrong either with our estimates of vacuum energy or
with our idea that energy is the source of gravity. The myriad
proposals to deal with the problem are given in the numerous
references of \cite{wei,car,pad}. Here this problem is related to
the structure of vacuum of QFT.

The idea here is that, like in quantum mechanics, one should
\textit{solve} for the ground state. Of course, one uses the
interaction picture, but that has the problem with the Haag's
theorem \cite{haag}. Also the success of the spontaneous symmetry
breaking idea \cite{weis,iz} seems to imply that vacuum is some kind
of condensate. The condensate states are generally not analytical in
the coupling constant (the prime example being BCS ground state
\cite{bcs}) so one can \emph{not} obtain those states
perturbativly from the non-interacting ground state.

In the next section the discussion of the need for the last
renormalization is given. In section three an explicit proposal for
a particular condensate ground state is given in the case of
$\varphi^4$ theory. In section four the perturbation theory with the
above ground state is discussed, while in section five nontrivial
vacuum solutions are presented. The last section is reserved for the
conclusion and outlook. An appendix is devoted to the condensate
state defined in terms of creation operators and the usual ground
state.
\section{The Last Renormalization}
\par In QFT one deals with the infinite energy of free oscillators by the
procedure called normal ordering, which forces vanishing of the
ground state energy to zeroth order in the coupling constant. That
one does by changing the order of operators in Hamiltonian and
Lagrangian \cite{bjo} \footnote {Here I am generally following the
notation of this textbook.} However, in calculating the propagator,
the inverse of the quadratic part of the Lagrangian, one uses
different ordering procedure since using the normal ordering would
lead to a vanishing propagator. These procedures, different
orderings for quadratic part of Lagrangian depending on what is
calculated, are standard in QFT. One pays the price in having the
perturbativly calculated correction to the vacuum energy infinite to
any order in the coupling constant. Note that the degree of the
divergence is $V\Lambda^4$, where V is normalization volume using
periodic boundary conditions while $\Lambda$ is the large momentum
cut-off. This degree is the same in the divergence of vacuum energy
removed by the normal ordering procedure and in the corrections to
vacuum energy in perturbative calculation. The divergence of vacuum
energy is indeed the most divergent quantity of all ultraviolet
divergent quantities in QFT. The hope is that if one can make
somehow this finite, all less divergent quantities (various
renormalization constants) which diagrammatically are sub-diagrams
of vacuum to vacuum diagrams, could be made finite too.

The relationship between the vacuum energy and ultraviolet
divergences in QFT was discussed by Allcock \cite{all}, and made
explicit by the supersymmetric theories which have vanishing vacuum
energy and milder ultraviolet divergences \cite{wbr,mar}. For
example, the Wess-Zumino model \cite{wz} is finite save for the wave
function renormalization \cite{gri}, while e.g. the $N=4$
supesymmetric Young-Mills is complectly finite \cite{soh}. For review
of divergences in supersymmetric theories see, e.g. \cite{pw}.

Here I consider the simple self interacting scalar $\phi^4$ field
theory. To make the theory finite it is not sufficient that the
vacuum energy be made finite. This is clear from the supersymmetric
theories which have the vanishing vacuum energy, many of which still
have divergences. Therefore it is not enough just to render the
connected vacuum to vacuum graphs, figure \ref{fig_2}, finite.
\FIGURE[h]{\epsfig{file=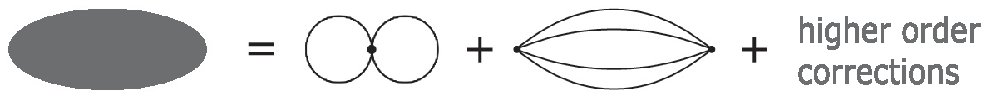,width=10cm}
        \caption{Connected vacuum to vacuum diagrams in
$\phi^4$ field theory giving perturbative corrections to vacuum
energy density}
	\label{fig_2}}
One also needs the functional derivative of this with respect to
propagator to be finite. That means if one has a functional of
propagators in k space, $F\left(\Delta(k_j)\right)$ given by
integrals over various k's one varies each propagator by
$\eta\delta(k_j-p)$ and subtracts unvaried expression and divides by
$\eta$ and finally performs $\eta\rightarrow 0$ limit. One
derivative applied to vacuum to vacuum diagram cuts a line of the
graph giving the propagator correction, while two derivatives cut two
lines giving scattering correction figure \ref{fig_3}, etc.
\FIGURE[h]{\epsfig{file=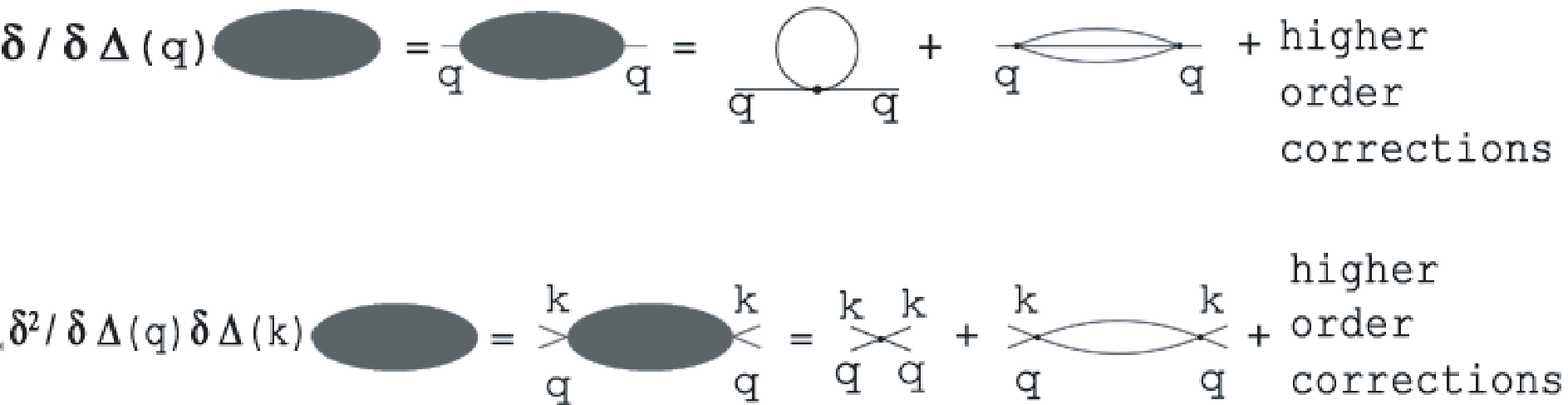,width=11cm}
        \caption{functional derivatives with respect to propagator}
	\label{fig_3}}
That is, functional derivatives with respect to the propagator cut
lines and one needs in addition to vacuum energy being finite, its
functional derivatives with respect to propagator to be finite. In
$\varphi^4$ theory one needs first four derivatives of the vacuum
energy with respect to the propagator to be finite to ensure the
finiteness of the theory. One can presumably write analogous
derivative relations for other field theories.

The whole spirit of renormalization is to equate a renormalized
quantity with an experimentally measured value. Since a finite
vacuum energy density have been measured \cite{per,per1,sch,reiss},
one should apply the renormalization procedure to the vacuum energy
too. One needs to see how one could implement what could be called
the last renormalization: \emph{renormalization of the vacuum
energy in such a way that all renormalization constants become
finite}. A proposal of what could be done, follows in
the next section.
\section{$\gamma$ Vacuum}
\par Consider the standard $\varphi^4$ field theory in four space-time
dimensions
\begin{eqnarray}
\label{lagdef} {\cal L} = \frac{1}{2}\frac{\partial\varphi}{\partial
x_\mu} \frac{\partial\varphi}{\partial x^\mu} -
\frac{1}{2}\:m^2\:\varphi^2 + \frac{1}{2}\:g\: \varphi^4 .
\end{eqnarray}
Define a condensate ground state by the relation:
\begin{equation}
\label{gvdef} \left(a^{\dagger}_k a_k +
\frac{\gamma}{2}\right)|v>_{\gamma}\:=\:0.
\end{equation}
Here, e.g. $a_k^{\dagger}$ is normalized creation operator (here I am
\textit{not} following the notation of Bjorken and Drell instead I am
following \cite{iz1}, with the exception of the covariant
normalization of states), with normalization
$\sqrt{V}a_k^{\dagger}=b_k^{\dagger}$ where
$[b_k,b_{k'}^{\dagger}]=\delta_{k,k'}V$ with box normalization used
and V normalization volume. In the continuum limit, the limit $V
\rightarrow \infty$ is performed. In this $\gamma$ is a parameter
that vanishes for normal ordering, while it is one to the lowest
order for natural ordering (meaning without any additional ordering,
see e.g. \cite{bd1}). The other values of $\gamma$ interpolate
between this cases. Since the ultraviolet divergences of free
oscillators and perturbative corrections are of the same type
($V\Lambda^4$), one can in principle adjust $\gamma$ so the energy
of free oscillators cancels the leading ultraviolet divergence of
the perturbative correction to vacuum energy to any order, all order
corrections to the vacuum energy having the same $V\Lambda^4$
dependance. Starting with the natural ordering and $\gamma=1$  and
changing $\gamma$ while keeping the natural ordering one gets free energy  proportional to $1-\gamma$.
With correction one expects
\begin{equation}
\label{ex} \gamma= 1+\mbox{const}\: g + \mbox{higher order
corrections}.
\end{equation}
Here g is the coupling constant and the value of $\gamma$ is
adjusted to cancel the leading $\Lambda^4$ divergence of the vacuum
energy. One hopes that canceling of next to leading singularities
leads to a relationships between mass and the coupling constant of
the theory, possibly involving renormalized (and experimentally
measured) vacuum energy. It would be shown later that the radiative
correction to the energy also depends on $1-\gamma$. The idea is
here that like in quantum mechanics, starting with an appropriate
initial state (which can be non-analytical in the coupling constant)
one calculates the corrections to the ground state order by order in
the coupling constant. Note that to the lowest order the two ground states with
different $\gamma$'s differ by an infinite energy and as a result
there is a superselection rule between words built upon those states
\cite{haag,nam}, the origin of the Haag's theorem. I shall discuss
an explicit realization of the $\gamma$ condensate shortly.

Examine now some properties of the $\gamma$ vacuum. It is clear that
the definition of $\gamma$ vacuum (\ref{gvdef}) introduces the
states of negative norm in the theory. Assuming the $\gamma$ vacuum
to be normalized one obtains from (\ref{gvdef})
\begin{equation}
\label{-1par} _{\gamma}<v|\left(a_k^{\dagger}a_k +
\frac{\gamma}{2}\right)|v>_{\gamma}=0 \quad \mbox{from which}\quad
\|a_k|v>_{\gamma}\|^2 = -\frac{\gamma}{2}.
\end{equation}
As will become clear from an explicit expression for the $\gamma$
vacuum one could have non vanishing states of an arbitrary number of
$a_k$ operators acting on the state, odd number of $a_k$'s giving
negative norm, as above. Note that negative norm states have been
both introduced \cite{dir} and criticized \cite{pau} relatively
early in the history of field theory, for a review see \cite{pei}.
\par As we will see shortly, the role of the negative norm states is to
provide high momentum subtraction in the theory, so perhaps their
unusual properties should not be a surprise. We have one well
established example of states doing substraction: the role of the
ghost states is to act like a negative degrees of freedom
\cite{fad, sw2}, designed to subtract the overcounting of the gauge
degrees of freedom. Those states do not satisfy the second Pauli
postulate \cite{pau1}, so that the fields do not commute for a space
like intervals, and that is certainly unusual. One can easily see
that the negative norm states are also states of negative energy,
and thus the states violate the first Pauli postulate. An
energy-parity symmetry with the negative energy ghost sector have
been discussed by Kaplan and Sundrum \cite{sun} in relation to
cosmological constant problem. Here one does not have energy-parity
symmetry: the negative energy states have negative norm for odd
number of particles, and it would be shown that the magnitude of the
norm of positive and negative energy states is different for
$\gamma\neq 1$.
\par With $\gamma$ vacuum one defines the number operator
of particles with momentum k by
\begin{equation}
\label{numop} N_k^{\gamma}= a_k^{\dagger}a_k +\frac{\gamma}{2}
\end{equation}
while the total number operator is
\begin{equation}
\label{tnumop} N=\sum_{k} \left(a_k^{\dagger}a_k +
\frac{\gamma}{2}\right).
\end{equation}
It is assumed that the operators $N$ and $N_k^{\gamma}$ are well
defined acting on states built by repeated application of
$a_k^{\dagger}$ or $a_p$ on $|v>_{\gamma}$. This one can use to show
that $a_k^{\dagger}|v>_{\gamma}$ represents one particle state by
\begin{equation}
\label{ops}
Na_k^{\dagger}\,|v>_{\gamma}\;=\;a_k^{\dagger}N\,|v>_{\gamma}+[N,a_k^{\dagger}]\:|v>_{\gamma}\;=\;
a_k^{\dagger}\,|v>_{\gamma}.
\end{equation}
Thus we can call $a_k^{\dagger}|v>_{\gamma}$ one particle state.
Analogously, one justifies calling the $a_k|v>_{\gamma}$ minus one
particle state.
\par One can use N operator to show vanishing of
$_{\gamma}<v|a_k^{\dagger}|v>_{\gamma}$ matrix element viz.
\begin{equation}
\label{mel}
_{\gamma}<v|a_k^{\dagger}|v>_{\gamma}\:=\:_{\gamma}<v|\left(Na_k^{\dagger}\right)|v>_{\gamma}\:=\:
{\gamma}<v|N a_k^{\dagger}|v>_{\gamma}\:=\:0
\end{equation}
which vanishes through action of the operator N on the left
$_{\gamma}<v|$. Analogously one shows vanishing of average value of
any odd number of $a_k^{\dagger}$ and $a_k's$. By the similar method (using $N_k^{\gamma}$) one ensures the vanishing of
$_{\gamma}<v|a_k^{\dagger}a_k'|v>_{\gamma}$ for $k\neq k'$.

One can use the the standard commutation relation between a's,
namely $\left[a_{k'},a^{\dagger}_k\right]=\delta_{k,k'}$, to
calculate the norm of a multiparticle state
$\left(a^{\dagger}_k\right)^n|v>_{\gamma}$. The result is
\begin{equation}
\label{norm}
\left\|\left(a^{\dagger}_k\right)^n|v>_{\gamma}\right\|^2=(n-\frac{\gamma}{2})(n-1-\frac{\gamma}{2})\dots(1-\frac{\gamma}{2})=
\prod_{j=1}^n(j-\frac{\gamma}{2}).
\end{equation}
Note that this expression tends to the usual $n!$ as $\gamma$ tends
to zero. Also note that this norm is decreasing with increasing
$\gamma$, vanishing for $\gamma = 2$. One can similarly calculate
the norms of the $\left(a_k\right)^n|v>_{\gamma}$ states, and the
result is
\begin{equation}
\label{norm-}
\left\|\left(a_k\right)^n|v>_{\gamma}\right\|^2=(-1)^n(\frac{\gamma}{2}+n-1)(\frac{\gamma}{2}+n-2)\dots\frac{\gamma}{2}=
\prod_{j=0}^{n-1}{-\left(\frac{\gamma}{2}+j\right)}.\qquad
\end{equation}
Again this vanishes as usual as $\gamma$ tends to zero. The
magnitude of this function is an increasing function of $\gamma$. Note
also that for $\gamma=1$ the norms of $|-n_k>_{\gamma}\:
\equiv\left(a_k\right)^n|v>_{\gamma}$ state and
$|n_k>_{\gamma}\:\equiv\left(a^{\dagger}_k\right)^n|v>_{\gamma}$
state are equal for even number of particles, while for an odd
number of particles they are equal in magnitude and different in
sign.

The matrix elements are very similar to usual. For any combination
of operators $a_k$ and $a_{k'}$ where $N\left(a_ka_{k'}\dots
a_{k_1}^{\dagger}a_{k_1'}^{\dagger}\dots\right)|v>_{\gamma}=q\left(a_ka_{k'}\dots
a_{k_1}^{\dagger}a_{k_1'}^{\dagger}\dots\right)|v>_{\gamma}$ and $q
\neq 0$ one can show that the scalar product with $_{\gamma}<v|$
vanishes, simply by
\begin{equation}
\label{me} _{\gamma}<v|\left(a_ka_{k'}\dots
a_{k_1}^{\dagger}a_{k_1'}^{\dagger}\dots \right)|v>_{\gamma}=
\frac{1}{q}\left(_{\gamma}<v|Na_ka_{k'}\dots
a_{k_1}^{\dagger}a_{k_1'}^{\dagger}\dots |v>_{\gamma}\right)=0.
\end{equation}
In particular that implies that the matrix element of odd number of
any kind of $a$'s vanishes as does
$_{\gamma}<v|a_k^{\dagger}a_{k'}|v>_{\gamma}$ for $k \neq k'$. The
later matrix element is non-vanishing only for $k=k'$. Thus some
simple matrix elements are
\begin{equation}
\label{metwo}
_{\gamma}<v|a_k^{\dagger}a_k'|v>_{\gamma}\:=\:-\frac{\gamma}{2}\delta_{k,k'}\qquad
_{\gamma}<v|a_ka_{k'}^{\dagger}|v>_{\gamma}\:=\:\left(1-\frac{\gamma}{2}\right)\delta_{k,k'}.
\end{equation}\\
For matrix elements of more operators, one obtains some terms in
addition to the standard ones, namely
\begin{eqnarray}
\label{mefour}
_{\gamma}<v|a_{k_1}a_{k_2}a_{k_3}^{\dagger}a_{k_4}^{\dagger}|v>_{\gamma}\:=\:
\left(\delta_{k_1,k_3}\delta_{k_2,k_4}+\delta_{k_1,k_4}\delta_{k_2,k_3}\right)\left(1-\frac{\gamma}{2}\right)^2
& + & \nonumber\\
\;+\;\delta_{k_1,k_2}\delta_{k_2,k_3}\delta_{k_3,k_4}\frac{\gamma}{2}\left(1-\frac{\gamma}{2}\right)\,.
\end{eqnarray}
The matrix elements for different order of operators can be easily
obtained by commutation. Note that the last term, that is different
then standard, becomes of the measure zero in the continuum limit.
Also it vanishes in the limit $\gamma$ tends to zero (or two). All
matrix element of more $a$'s have analogous terms, always
proportional to $\frac{\gamma}{2}\left(1-\frac{\gamma}{2}\right)$.
Let us just quote the result for six operators
\begin{eqnarray}
\label{mesix} \nonumber
_{\gamma}<v|a_{k_1}a_{k_2}a_{k_3}a_{k_4}^{\dagger}a_{k_5}^{\dagger}a_{k_6}^{\dagger}|v>_{\gamma}
& = &
\left(1-\frac{\gamma}{2}\right)^3\left(\delta_{1,4}\delta_{2,5}\delta_{3,6}+\mbox{perm
1,2,3}\right) +  \\  & + &
\frac{\gamma}{2}\left(1-\frac{\gamma}{2}\right)^2\times
\left(\delta_{1,4}\delta_{2,3,5,6}+ \mbox{combinations} \right)+\\
\nonumber
+\gamma\left(1-\frac{\gamma}{2}\right)\left(\gamma-1\right)\delta_{1,2,3,4,5,6}.
\end{eqnarray}\\
Here, for example $\delta_{2,3,5,6}$ stands for $\delta_{k_2,\:k_3}\delta_{k_3,\:k_5}\delta_{k_5,\:k_6}$ and analogous notation for other indices. Note that all nonstandard terms are of measure zero in the continuum limit and also vanish as $\gamma$ tends to zero. Again one can obtain the matrix elements for other orders of operators simply by commutation.
\par Let us now give an explicit expression for $|v>_{\gamma}$ in terms od the standard vacuum $|0>$\,,
\begin{equation}
\label{defv} |v>_{\gamma}\:=\:
\prod_{k_1}\left(\int_0^{\infty}e^{-(a_{k_1}^{\dagger}a_{k_1}^{\dagger}t_{k_1})}
t_{k_1}^{\frac{\gamma}{4}-1} dt_{k_1}\right) |0>.
\end{equation}
Since this is a tensor product of different k's one can have various C(k) inserted in the product, I am using one as the simplest.
Note that the limits for $\gamma$ are
\begin{equation}
\label{gamlim} 0<\gamma<2.
\end{equation}
 Parameter $\gamma$ starting from one to the lowest order should not change through radiative corrections by one or more.
The derivations of the above equations are given in the appendix.
\section{Perturbation theory with $\gamma$ Vacuum}

\par The first step in developing perturbation theory is to calculate the propagator. Defining
\begin{eqnarray}
\label{gamapro}
 i\Delta_{\gamma}(x) \equiv
_{\gamma}<v|T(\varphi(x)\varphi(0)|v>_{\gamma}& = &\nonumber \\
\nonumber = \;i\left(1-\frac{\gamma}{2}\right)\int
\frac{d^4k}{(2\pi)^4}\frac{1}{k^2-m^2+i\epsilon}e^{-ikx}+
i\frac{\gamma}{2}\int\frac{d^4k}{(2\pi)^4}\frac{1}{k^2-m^2-i\epsilon}e^{-ikx} & = &\\
=\; i\int\frac{d^4k}{(2\pi)^4}\frac{1}{k^2-m^2+i\epsilon}e^{-ikx}-
\frac{\gamma}{2}\int\frac{d^4k}{(2\pi)^3}\delta\left(k^2-m^2\right)e^{-ikx} & =& \\
\nonumber = \; i\Delta_F(x) - \frac{\gamma}{2}\Delta_1(x).\qquad
\end{eqnarray}
\par Here I am using the notation of Bjorken and Drell, T stands for time
ordered product, $\Delta_F(x)$ for Feynman propagator, while
$\Delta_1(x)$ is even solution for the homogenous Klein-Gordon
equation. The above equation is derived by expanding the field
$\varphi$ in terms of creation and annihilation operators and using
the matrix elements (\ref{metwo}). We see here that the change of the
usual Feynman propagator does not effect an external (on mass shell)
line. The change of the propagator comes from two sources: first is
the change in the normalization of states, multiplying the usual
propagator by  $\left(1-\frac{\gamma}{2}\right)$, second is the
effect of negative norm states, propagating with the acausal
propagator characterized with $-i\epsilon$ prescription in going
around poles, the opposite sign to the usual Feynman prescription.
This makes the difference in continuation to the Euclidean $k_0$,
resulting in different signs which leads to some cancelations. In
going to Euclidean $k_0$ one starts with expressions of the form
$\int d^3k \oint f\left(k_0,\vec k\right) dk_0 = 0$ where $\oint$ is
integral over closed path in a complex $k_0$, contour being such to avoid
singularities. For contours see the enclosed figure.
\FIGURE[h]{\epsfig{file=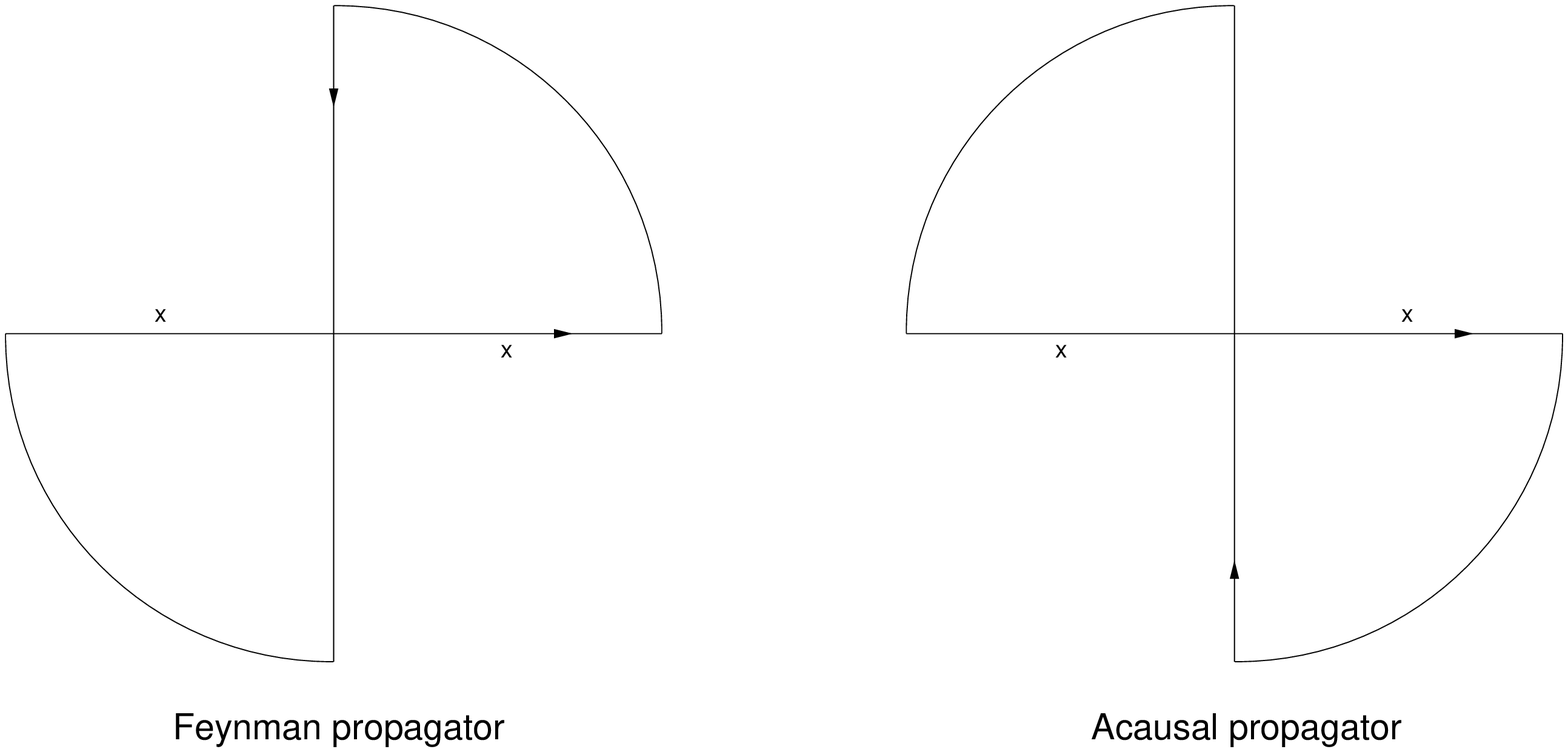,width=10cm}
        \caption{The contours for Feynman and for acausal propagator}
\label{fig_4}}
The contours for Feynman propagator and for acausal propagator for negative norm states are different and where for Feynman propagator one gets
\begin{equation}
\label{Euclf}
\int d^3k \int_{-\infty}^{\infty}f\left(k_0,\vec k\right) dk_0 \:=\: i \int d^3k \int_{-\infty}^{\infty}f\left(iy,\vec k\right) dy
\end{equation}

for the acausal propagator one gets

\begin{equation}
\label{Eucla} \int d^3k \int_{-\infty}^{\infty}f\left(k_0,\vec
k\right) dk_0 \:=\: -i \int d^3k
\int_{-\infty}^{\infty}f\left(-iy,\vec k\right) dy.
\end{equation} Here y plays the role of Euclidean $k_0$.

Consider now the second order correction to the scattering
amplitude, see figure \ref{fig_5}.
\FIGURE[h]{\epsfig{file=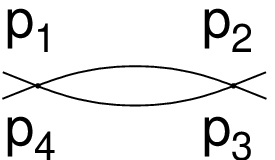,width=3cm}
\label{fig_5}
        \caption{correction to scattering amplitude}}
In this one can follow the textbook derivation \cite{bro}
and use dimensional regularization. Consider the integral
\begin{equation}
\label{gamacor} F_{\gamma \mbox{reg}}\left(p^2\right) =
\int\frac{d^nk}{(2\pi)^n} \Delta_{\gamma}(k)\Delta_{\gamma}(k+p).
\end{equation}
Here p represent different momentum indices (all three Mandelstam variables s, t, u see \cite{bro}, and could be assumed space-like and if needed continued to time-like). Since the integral is a Lorentz-invariant function of $p^2$ for space-like p one can assume $p_0 = 0$. One can then continue the integrals in $k_0$ into complex plane with appropriate contours.
Using in the above equation definition of $\gamma$ propagator (\ref{gamapro}) one gets the equation for F consisting of four integrals
\begin{eqnarray}
\label{corscat}
F_1 & = & \nonumber \left(1-\frac{\gamma}{2}\right)^2\int
\frac{d^nk}{(2\pi)^n}\frac{1}{k^2-m^2+i\epsilon}
\frac{1}{(k+p)^2-m^2+i\epsilon}\\
F_2 & = & \left(1-\frac{\gamma}{2}\right)\left(\frac{\gamma}{2}\right)\int
\frac{d^nk}{(2\pi)^n}\frac{1}{k^2-m^2+i\epsilon}
\frac{1}{(k+p)^2-m^2-i\epsilon}\\
F_3 & = & \nonumber \left(1-\frac{\gamma}{2}\right)\left(\frac{\gamma}{2}\right)\int
\frac{d^nk}{(2\pi)^n}\frac{1}{k^2-m^2-i\epsilon}
\frac{1}{(k+p)^2-m^2+i\epsilon}\\
F_4 & = & \nonumber \left(\frac{\gamma}{2}\right)^2\int
\frac{d^nk}{(2\pi)^n}\frac{1}{k^2-m^2-i\epsilon}
\frac{1}{(k+p)^2-m^2-i\epsilon}.
\end{eqnarray}
The integrals one and four we can deal with by standard methods \cite {bro}, only the rotation to Euclidean momenta go by different contours. Lets add the integrals two and three
\begin{eqnarray}
 F_2 + F_3  =  \nonumber
 \label{I2I3}
\left(1-\frac{\gamma}{2}\right)\left(\frac{\gamma}{2}\right)\int
\frac{d^nk}{(2\pi)^n}[\frac{1}{k^2-m^2+i\epsilon}\frac{1}{(k+p)^2-m^2-i\epsilon}]+\\
+ \nonumber \left(1-\frac{\gamma}{2}\right)\left(\frac{\gamma}{2}\right)\int\frac{d^nk}{(2\pi)^n}
[\frac{1}{k^2-m^2-i\epsilon}\frac{1}{(k+p)^2-m^2+i\epsilon}]  =  \\
=  \left(1-\frac{\gamma}{2}\right)\left(\frac{\gamma}{2}\right)\int
\frac{d^nk}{(2\pi)^n}[\frac{1}{k^2-m^2+i\epsilon}\frac{1}{(k+p)^2-m^2+i\epsilon}]&&+\nonumber \\
+ \left(1-\frac{\gamma}{2}\right)\left(\frac{\gamma}{2}\right)\int
\frac{d^nk}{(2\pi)^n}[\frac{1}{k^2-m^2-i\epsilon}\frac{1}{(k+p)^2-m^2-i\epsilon}] + \\ +
\nonumber\left(1-\frac{\gamma}{2}\right)\left(\frac{\gamma}{2}\right)
4\epsilon^2 I_1 \:,
\end{eqnarray}
where
\begin{equation}
\label{I1}
I_1=\int\frac{d^nk}{(2\pi)^n}\frac{1}{[k^2-m^2]^2-\epsilon^2}\frac{1}{[(k+p)^2-m^2]^2-\epsilon^2}\:.
\end{equation}
For nonzero $p$ the last term of \ref{I2I3} vanishes as $\epsilon$ when $\epsilon$ tends to zero while the other two terms can be added to $F_1$ and $F_4$ respectively. One can calculate the explicit expressions following Brown, \cite {bro}, the result for the amplitude is
\begin{equation}
\label{amplitude}
F_{\gamma \mbox{reg}}\left(p^2\right) = i(1-\gamma) \Gamma(2-n/2)\frac{\mu^{n-4}}{(2\pi)^2} \int_0^1\left(\frac{m^2+\alpha(1-\alpha)p^2}{4 \pi \mu^2}\right)^{n/2-2}d\alpha\:,
\end{equation}
where $\mu$ is an arbitrary mass scale.
The key of the matter is that the result is proportional to $i(1-\gamma)$. That means that starting with $\gamma=1$ in lowest approximation, the radiative correction vanishes and the loop corrections are zero. Moreover, since other more involved corrections are obtained by integrating this expression, those vanish too.
\par One has to have in mind that we are using neither normal ordering nor Wick's theorem here and one has to consider corrections to propagator and vacuum energy with the lines starting and finishing at the same vertex, see figure.
\FIGURE[htb]{\epsfig{file=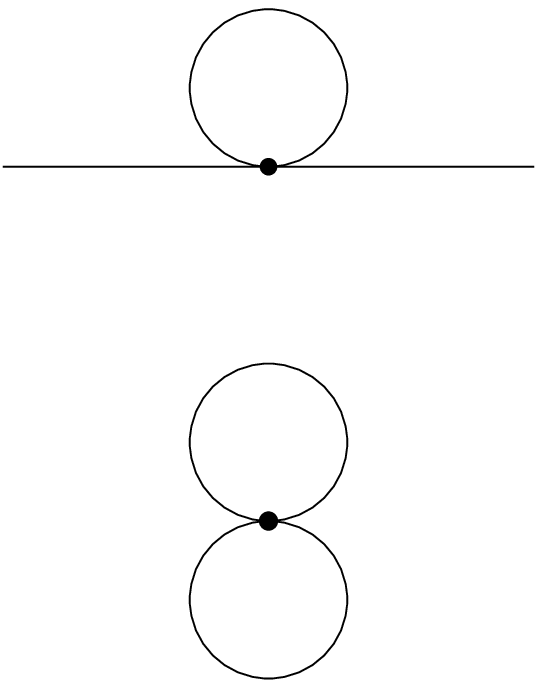,height=3cm}
        \caption{Corrections for the propagator and vacuum energy}
\label{fig_6}}
 Calculating directly (strictly speaking one does not use time ordering in these lines, but the results with, or without, time ordering are the same) one obtains the propagator correction
\begin{equation}
\label{propcor}
i(1-\gamma)\int\frac{d^nk_E}{(2\pi)^n}\frac{1}{k_E^2+m^2}\,.
\end{equation}
Here $k_E$ is Euclidean momentum and this vanishes for $\gamma=1$.
It is perhaps simpler to show this in x space, instead in k space
\footnote {using the notation and formulae from Bjorken and Drell
field theory, p 388.}. One has $$i\Delta_{\gamma}(x)=
i\Delta_F(x)-\frac{\gamma}{2}\Delta_1(x)=\frac{1}{2}(1-\gamma)\Delta_1(x)
+i\frac{\epsilon(t)}{2}\Delta(x).$$ When $\gamma=1$ the first part
is zero, while the function $\Delta(x)$ is odd and vanishes at
$x=0$. So the lines starting and finishing at the same point do not contribute.
\par One here obtains an unusual result: The radiative corrections for the $\varphi^4$ field theory vanish with this formulation. This theory is in some sense trivial. The arguments for triviality of the $\varphi^4$ theory are well known \cite{soc, fer, sim, call, wil, lan, pom}, \cite {wil1, froh, freed, lang, huang, kuti, kuti1, but}, both analytically and through computer simulation, however the arguments here are perturbative and thus different. In contrast the standard perturbative treatment \cite{bro} certainly gives nontrivial correction to the scattering amplitude. Here one obtains perturbatively the vanishing of the radiative corrections; the $\gamma$ vacuum brings agreement between perturbative calculation and other methods.
\section{Nontrivial $\gamma$}
Now let us investigate the vacuum energy for $\gamma \neq 1$. In that case one has, to lowest order, vacuum energy proportional to $1-\gamma$. To that order the vacuum energy is given by
\begin{eqnarray}
\label{vezero}
E_0 & = & \nonumber \frac{1}{2}(1-\gamma)\int \frac{d^3k}{(2\pi)^3}\sqrt{k^2+m^2} = \\
& = & \frac{(1-\gamma)}{(2\pi)^2}\int_0^{\Lambda}k^2\sqrt{k^2+m^2}dk = \\
& = & \nonumber \frac {(1-\gamma)\Lambda^4}{8(2\pi)^2}\left[\sqrt{1+a^2}(2+a^2)+a^4\ln\frac{a}{1+\sqrt{1+a^2}}\right]. \end{eqnarray}

Here $a \equiv \frac{m}{\Lambda}$. All logarithmic divergences are represented by $\ln{a}$.
Note that here we are using three dimensional high momentum cut-off $\Lambda$, four dimensional is not convenient since we are on the mass hyperboloid. With three dimensional high momentum cut-off, one expects some Lorentz invariance violation close to the cut-off momentum. That violation was not seen \cite{kos, matt, jac}, which means that cut-off is high, in the parlance of this model, $a$ is small.
\par One can now calculate the correction to the vacuum energy to order $g$, using the vacuum to vacuum diagram from the figure \ref{fig_6}, just the average value (using $|v>_{\gamma}$) of the interaction term. After some calculation one obtains
\begin{equation}
\label{veone} E_1 = \frac {3g(1-\gamma)^2}{2^3\pi^4}I^2 \quad\mbox{
where }\quad I = \int_0^{\Lambda}\frac{k^2}{2\sqrt{k^2+m^2}}\:dk .
\end{equation}
That integral can be calculated from the previous one $E_0$ integral \ref{vezero} by differentiation with respect to $m^2$. Calculating this one obtains
\begin{equation}
\label{veonec}
E_1 \: = \: \frac{3g(1-\gamma)^2}{(2\pi)^4}\: \frac{\Lambda^4}{8}\,f_2(a)
\end{equation}
 where
\begin{equation}
\label{f2}
\textstyle
 f_2(a) \equiv
\frac{1}{1+a^2}\left[1+3a^2/4-\frac{a^4}{4(1+\sqrt{1+a^2})}+\frac{a^2}{4}\sqrt{1+a^2}+
a^2\sqrt{1+a^2}\ln{\frac{a}{1+\sqrt{1+a^2}}}\right]^2.
\end{equation}
Note that $f_2(a)$ is nonnegative, and $f_2(0)=1$.
One then obtains
\begin{equation}
\label{energycorr} E_0+E_1 = \frac{\Lambda^4}{8}
\left[\frac{1-\gamma}{(2\pi)^2}f_1(a) +
\frac{3g(1-\gamma)^2}{(2\pi)^4}f_2(a)\right].
\end{equation} \\
Here $f_2(a)$ is defined above, while $f_1(a)$ is given by
\begin{equation}
\label{def f_1} f_1(a) \equiv
\left[\sqrt{1+a^2}(2+a^2)+a^4\ln\frac{a}{1+\sqrt{1+a^2}}\right].
\end{equation}
Note that $f_1(0)=2$. Obviously, $\gamma=1$ does make the $\Lambda^4$ term vanish, however this is not the only possibility. The other possibility (to the first order in g) is given by
\begin{equation}
\label{nontr} f_1(a)+\frac{3g(1-\gamma)}{(2\pi)^2}f_2(a)=0.
\end{equation} Solving for $\gamma$ one obtains
\begin{equation}
\label{sol-gamma} \gamma = 1 +
\frac{(2\pi)^2}{3g}\frac{f_1(a)}{f_2(a)}.
\end{equation}
Note that this is \emph{non-analytical} in $g$ at $g=0$, having dependence $\displaystyle \frac{1}{g}$\,. Since this $\gamma \neq 1$, one gets nonzero scattering amplitude correction equation (\ref{amplitude}) different from the trivial case $\gamma = 1$.
 Also we have the limits for the value of $\gamma,\; 0<\gamma<2\;$, putting that in the above expression for $\gamma$ one obtains
\begin{equation}
\label{limg} \frac{g}{2} > \frac{(2\pi)^2}{3}\frac{f_1(a)}{2f_2(a)}
\sim \frac{(2\pi)^2}{3}\left(1-2a^2\ln(a/2)\right).
\end{equation}
The last behavior is for small $a$.
That means that there is a possibility of phase transition in this model at $g/2$ greater then $\frac{(2\pi)^2}{3}$, although the energy is not lower then $\gamma =1$ case. The theory should be checked in computer simulations in this range of the coupling for any sign of phase transition.
\par We proceeded here assuming the the coefficient $\Lambda^4$ of the vacuum energy vanishes. However, since the expression for vacuum energy equation (\ref{energycorr}) is proportional to $\Lambda^4$, in that approximation, the whole vacuum energy vanishes. Now lets check what happens if we presume finite vacuum energy density, as evidenced by observation \cite{per, per1, sch, reiss}. Using the expression for the vacuum energy to first order in the coupling (\ref{energycorr}), one obtains
\begin{equation}
\label{eneq} E_0+E_1 = \frac{\Lambda^4}{8}\frac{1-\gamma}{(2\pi)^2}
\left[f_1(a) + \frac{3g(1-\gamma)}{(2\pi)^2}f_2(a)\right] = E_m .
\end{equation}\\
Here one can take $E_m$ a measured vacuum energy density (of the order of $(10^{-3}eV)^4$) or some other value in a model calculation. The $f_1(a)$ and $f_2(a)$ are defined previously, equations (\ref{def f_1}), (\ref{f2}). By this equation we are changing problem from the one of cancelation, to the one of fine tuning the constants so equation (\ref{eneq}) is satisfied. Defining
$ X \equiv \frac{1-\gamma}{(2\pi)^2}$ one obtains a quadratic equation for $X$
\begin{equation}
\label{eqx} Xf_1(a) + 3gX^2f_2(a) = \epsilon \qquad \epsilon \equiv
\frac{8E_m}{\Lambda^4} .
\end{equation}
Two solutions for $X$ give two solutions for $\gamma$. One solution bit smaller then one (for positive $\epsilon$) a bit greater then one for negative $\epsilon$, and analytical function of $g$. The other greater then one, non-analytical function of $g$. Those are given by
\begin{equation}
\label{gamsol} \gamma_{(1,2)} = 1 + \frac{(2\pi)^2 f_1(a)}{6g
f_2(a)}\times \left(1 \mp
\sqrt{1+12g\epsilon\frac{f_2(a)}{(f_1(a))^2}}\right)\,.
\end{equation}
Here $\epsilon$ is very small and expanding the above solution in series one obtains
\begin{equation}
\gamma_1 = 1 - \frac{(2\pi)^2 \epsilon}{f_1(a)} + 0(\epsilon^2)\qquad
\gamma_1 = 1 + \frac{(2\pi)^2 f_1(a)}{3 g f_2(a)} + 0(\epsilon)
\label{gama12}\:.
\end{equation}
Using the limits on $\gamma$,
and expanding the square root for small $\epsilon$ one obtains the limits
\begin{equation}
\label{limep}
\mid \epsilon/f_1(a) \mid < \frac{1}{(2\pi)^2}
\end{equation}
for the first sign and the same limit as previously, equation (\ref{limg}), for the second sign. Therefore by fine tuning the constants, one can obtain the finite energy density within this model. Given the smallness of $\epsilon$ the first sign is very close to $\gamma=1$ case with no radiative corrections. Again the theory has the other sector with nontrivial radiative corrections and specific range of the coupling. Both $\gamma < 1$  (for positive $\epsilon$) function analytical in g, and $\gamma > 1$, function non-analytical in g are possible.
\section{Conclusion and Outlook}
\par It is shown here that handling of the worst $(\Lambda^4)$ singularity in the vacuum energy is possible with the condensate defined by (\ref{gvdef}), starting from natural instead of normal ordering. Triviality of the $\varphi^4$ theory is explained perturbatively for $\gamma = 1$. However for $\gamma \neq 1$ there is a possibility of a non-trivial state non analytical in the coupling constant equation (\ref{sol-gamma}) and one solution of equation (\ref{gamsol}), and those can not be obtained by the standard perturbation theory.
\par The moral of this story is the existence of negative energy sector of QFT that has a subtractive role. Out of four integrals needed to bring perturbative calculation in agreement with non-perturbative methods (equation (\ref{corscat})) the standard method recognisees only one; that leads to disagreement between the methods which was obviated in this paper. The corollary to this moral is that no symmetry is needed to ameliorate the ultraviolet divergences of QFT, what \emph{is needed is the proper solution for the ground state.}
\par There are, however, some unanswered questions. First, what about next $(\Lambda^2)$ singularity that appears in the propagator, or mass, correction for $\gamma \neq 1$? If we keep the standard Feynman diagram calculation there is nothing to cancel the divergent term given in equation (\ref{propcor}) and one would have to resort to the standard mass correction counter term. However, the propagator correction is the functional derivative with respect to propagator of the vacuum energy (that holds for all orders of diagrams) and one can take that as the method of defining the propagator corrections. In that case one also has to take the functional derivative of the zero point energy (\ref{vezero}) with respect to the propagator, $ \delta E_0/\delta\Delta(p)$. The first term is just function \footnote {one expects $(p^2-m^2)^2$ term but making Hamiltonian in theories with higher derivatives is involved \cite{pai,dur} and I will not calculate it here} of p while the second term is is obtained by taking the derivative of $E_0$ with respect to $m^2$ and then the functional derivative of $m^2$ with respect to $\Delta(p)$. To first order in $g$ the derivative of $m^2$ with respect to $\Delta(p)$ is just the coupling constant and one has the exactly the structure of the term needed to cancel the propagator infinity. Analogously by taking second derivative of $E_0$ with respect to $\Delta(p)$ one obtains the logarithmically divergent term needed to regularize the correction to scattering. Calculating in this manner may lead to cancelations.
\par Of course, getting the finiteness of a field theory in theories with unusual states is nothing new (see e.g. \cite{Lee, Ar, Gl}), what is new is the motivation for negative norm states, coming from the specific condensate vacuum. As opposed to other papers the hermiticity of the Hamiltonian and standard commutation relations are kept here (which means the standard microscopic causality \cite{bjo5}), the price payed is violating both the positivity of energy (the first Pauli postulate) and positivity of norm. Note however that with Lorentz invariant normalization the added term $2E$ makes the norm positive.
\par Second, the unitarity of the theory with these states is not discussed, for an earlier example of discussion of unitarity of theories with a non-positive norm see \cite{nak}. In my opinion, those "renormalizaton ghost" states violating the first Pauli postulate are necessary for the complete set of states, and take full part in resolution of unity.  Those states are propagating forward in time with negative energy and it is not clear to me are there any other effects of those states besides the high momentum substraction. Given that those are negative energy states perhaps the states are tachyonic although in the usual formulation \cite{fei} tachyonic field has different commutation rules. Physics of negative energy states, as well as physics of vacuum structure, still has to be worked out.
 \par As outlook lets discuss (speculative) possibilities for further research. The key here is to introduce the fermions: for fermions the sign of the vacuum energy is opposite to that of bosons and that helps regularization of the vacuum energy. Note that for fermions the negative energy states describe antiparticles, and what one can glimpse from expressions for norms for the positive energy states (\ref{norm}), \footnote {having in mind that fermions have opposite sign of vacuum energy} decreasing with $\gamma$, and negative energy states (\ref{norm-}) increasing with $\gamma$ is the possible reason for the observed particle-antiparticle asymmetry: \emph{the situation is not symmetrical since vacuum energy density is different from zero , and $\gamma \neq 1$}. What one has is akin to energy parity symmetry \cite{sun} (with negative energy sector having negative norm and subtractive effect) softly broken by the nonzero, albeit small, vacuum energy density.
\par Presumably the higher order corrections makes an equation of the higher order then quadratic, which leaves the room for a richer vacuum structure, several solutions for $\gamma$ describing different vacua of the same energy\footnote{this is important since it avoids the superselection rule between the states built over these states} and possibly the different generations of particles. The same field with non-trivial vacuum structure  may, perhaps, describe generations.
\par If these speculations are valid, the study of the structure of vacuum is worthwhile; that should not be a surprise since for Euclidean field theory what one studies is like a statistical sum in statistical physics, a quantity of prime importance. What may be hiding in the vacuum structure is the information about particle spectrum. All this has to be extended to systems of interacting bosons and fermions, and appropriate symmetry breaking condensate searched for while protecting (to a good approximation) the Lorentz invariance. The finiteness of the vacuum energy density should be a requirement given the observed small vacuum energy density. In that one would be helped by the opposite signs of vacuum energies of bosons and fermions. So in addition to (and perhaps concurrently with) renormalizability, the existence of the appropriate vacuum state(s) may be the condition for a realistic field theory. If one is able to make the theory finite in the process, the question of Weisskopf \cite{we} on the soundness of renormalization procedure will be answered affirmatively. The measure of success of such a program would be the decreasing of the bewildering number of constants describing the standard model of particle physics  \cite{teg}.
 \acknowledgments
In the early phase of this work the author enjoyed the hospitality of Dipartimento di Scienze Fisiche, Universit\`{a} di Cagliari and INFN, Sezione di Cagliari, Italy. Author is grateful to the Institutions for the hospitality and to Alberto Devoto for many useful discussions. The access to libraries of the Department of Physics, University of Belgrade and Institute of Physics, Zemun, were very beneficial. Author is grateful to the Metrology Group of the Laboratory 011 of The "Vin\v ca" Institute of Nuclear Sciences for the use of their Mathematica program. Milutin Blagojevi\'c, Zoran Ivi\'c and Zoran Popovi\'c were helpful in preparation of the manuscript.
\appendix
\section{Standard Vacuum and $\gamma$ Vacuum}
Here the relation between the standard field theory vacuum and the condensate $\gamma$ vacuum is discussed. Act with $a_k^{\dagger}a_k$ on the expression for the $\gamma$ condensate state, (\ref{defv})
\begin{eqnarray}
\label{derv}
\nonumber a_k^{\dagger}a_k |x> \:\equiv \:a_k^{\dagger}a_k \prod_{k_1}\left(\int_0^{\infty}e^{-(a_{k_1}^{\dagger}a_{k_1}^{\dagger}t_{k_1})} t_{k_1}^{\beta - 1} dt_{k_1}\right)|0> \:=\\
\:=\:a_k^{\dagger}a_k\left(\int_0^{\infty}e^{-(a_{k}^{\dagger}a_{k}^{\dagger}t_k)}t_k^{\beta
- 1}dt_k\right)|0_k> \prod_{k_1 \neq
k}\left(\int_0^{\infty}e^{-(a_{k_1}^{\dagger}a_{k_1}^{\dagger}t_{k_1})}
t_{k_1}^{\beta - 1}dt_{k_1}\right)|0>.
\end{eqnarray}
Here $|x> $ is written instead of $|v>_{\gamma}$ until the equality of the two is proven, and ditto $\beta-1$ instead of $\frac{\gamma}{4}-1$.
Since we are dealing with the tensor product of expressions for the different k's, one can separate
\begin{eqnarray}
\nonumber \label{derkv}
a_k^{\dagger}a_k\left(\int_0^{\infty}e^{-(a_{k}^{\dagger}a_{k}^{\dagger}t_k)}t^{\beta - 1}_k dt_k\right)|0_k> & = & \\
= \; a_k^{\dagger} \int_0^{\infty} \left[a_k, e^{-(a_{k}^{\dagger}a_{k}^{\dagger}t)}\right] t^{\beta - 1} dt |0> & = & \\
\nonumber = \; 2 \int_0^{\infty} - a_k^{\dagger}a_k^{\dagger}
e^{-(a_{k}^{\dagger}a_{k}^{\dagger}t)}t^{\beta} dt |0_k> \,,
\end{eqnarray}
using partial integration with

 $dv  = - a_k^{\dagger}a_k^{\dagger} e^{-(a_{k}^{\dagger}a_{k}^{\dagger}t)} dt\qquad
\mbox{and} \qquad t^{\beta} = u $ \qquad one obtains for the above expression
\begin{eqnarray}
\label{part}
 - 2 \beta \int_0^{\infty}e^{-({a_k}^{\dagger}a_{k}^{\dagger}t)}t^{\beta - 1} dt |0_k> + 2 t^{\beta}e^{-(a_{k}^{\dagger}a_{k}^{\dagger}t)} |0_k> |_0^{\infty}.
\end{eqnarray}
That means if
\begin{equation}
\label{cond}
2 t^{\beta}e^{-{a_k}^{\dagger}a_{k}^{\dagger}t} |0_k> |_0^{\infty}\: =\: 0
\end{equation}
then
\begin{equation}
a_k^{\dagger}a_k |x> \:= \:-2\beta |x>
\end{equation}
or $\gamma\: = \:4\beta$ , $|x>\: = \:|v>_{4\beta}$. Obviously for $\beta > 0$ the expression (\ref{cond})
vanishes at $t\:=\:0$. To check what happens at $t\rightarrow\infty$ we need to check
\begin{eqnarray}
\parallel e^{-a_{k}^{\dagger}a_{k}^{\dagger}\, t }|0_k>
\parallel^2 & = & <0|e^{-a_{k}a_{k}\,t}e^{-a_{k}^{\dagger}a_{k}^{\dagger}\,t}|0>
\end{eqnarray}

as \quad $t\rightarrow\infty$.

\par Here we use the closed algebra of the dynamical group of harmonic oscillator \cite{wy} given by the elements
\begin{eqnarray}
\label{groupel} A = \frac {a_k^2}{2\sqrt{2}} \qquad B =\frac
{a_k^{2{\dagger}}}{2\sqrt{2}} \qquad  C  = \frac {\left(a_k
a_{k}^{\dagger} + a_{k}^{\dagger} a_k \right)} {4}.
\end{eqnarray}
The commutation relations are
\begin{equation}
\label{comrel}
 [A,B] = C \qquad [A,C] = A \qquad [B,C] = -B .
\end{equation}
Defining $$ x \equiv 2\sqrt{2}t$$ we write
\begin{equation}
\label{defom} \Omega(x) \equiv e^{-Ax} e^{-Bx} = e^{-B\beta(x)}
e^{-C\delta(x)} e^{-A\alpha(x)}.
\end{equation}
For $t \rightarrow \infty$ we are interested in behavior of $\delta(x), \mbox{as}\; x \rightarrow \infty$.
Differentiating (\ref{defom}) with respect to x, one obtains using repetitive commutation
\begin{eqnarray}
\label{difom}
-\left[\left(1+x^2/2\right)A + B - x C \right] \Omega(x) = \\
= \quad - \left[\alpha^{\prime} e^{\delta}A + \left(\beta^{\prime} + \delta^{\prime} \beta + \alpha^{\prime} e^{\delta} \frac{\beta^2}{2} \right)B + \left(\delta^{\prime} + \alpha^{\prime} e^{\delta}\beta \right)C \right] \Omega(x)
\end{eqnarray}.

Comparing the left and right sides of the above formula one obtains a system of differential equations for $\alpha, \beta$, and $\delta$. The initial conditions are $\alpha(0)=0,\quad \beta(0)=0,\quad \delta(0)=0$. The equations could be written as
\begin{eqnarray}
\label{difeks}
\nonumber 1+x^2/2 = \alpha^{\prime} e^{\delta} \\
1 = \beta^{\prime} + \delta^{\prime} \beta + (1+x^2/2)\frac{\beta^2}{2}\\
\nonumber - x = \delta^{\prime} + (1+x^2/2)\beta .
\end{eqnarray}
One can show the asymptotic behavior for large x (and large t) is
\begin{eqnarray}
\label{asbih}
\nonumber \beta \sim \frac{-2}{x} - \frac{4}{x^3} + 0\left(x^{-5}\right)\\
\mbox{and}\qquad \delta^{\prime} \sim \frac{4}{x} + 0\left(x^{-3}\right)\qquad \mbox{leading to}\\
\nonumber \delta \sim 4 ln\left(const\;{x}\right) +
0\left(x^{-2}ln{x}\right).
\end{eqnarray}
That enables us to conclude
\begin{equation}
<0| e^{-B\beta}\; e^{-C\delta}\; e^{-A\alpha} |0> \quad = \quad
<0|e^{-C\delta}|0> \quad = \quad \frac{1}{e^{\delta/4}} \sim
\frac{1}{x}\quad \mbox{for} \quad x \rightarrow \infty \,.
\end{equation}
That could be used for
\begin{equation}
\label{asnorm} { \parallel {t^{\beta} e ^{a_k^{\dagger2}t}
\parallel^2}} \sim  \frac{t^{2\beta}}{t} \quad
\end{equation}
as \quad $t
\rightarrow \infty$.

  Thus we need $ 0<\beta<1/2$ which gives
\begin{equation}
\label{limgam}
0< \gamma < 2 \;.
\end{equation}
Finally let us calculate the norm of $|v>_{\gamma}$ state. Define
\begin{equation}
\label{defomxy} \Omega(x,y) = e^{-Ax} e^{-By} = e^{-\beta(x,y)}
e^{-C\delta(x,y)} e^{-A\alpha(x,y)},
\end{equation} where A, B, C are the previously defined operators. Differentiating with respect to x and y and using methods analogous to already shown one can show that
$\delta = 2 ln \left(1-xy/2\right)$. That implies
\begin{equation}
\label{normom} <0|e^{-Ax}e^{-Bx}|0>\: = \:<0|e^{-C\delta}|0>\: =\:
<0|e^{-\delta/4}|0> \:=\: <0|0> \frac{1}{\sqrt{1-xy/2}} .
\end{equation}
Therefore
\begin{equation}
\label{normv}
_{\gamma}<v|v>_{\gamma} \:= \: \prod_k
\int_0^{\infty}\int_0^{\infty} (t\tau)^{\gamma/4 - 1}
\frac{1}{\sqrt{1-4t\tau}}\:dtd\tau <0|0>_k.
\end{equation} The integral
\begin{equation}
\int_0^{\infty}\int_0^{\infty} (t\tau)^{\gamma/4 - 1} \frac{1}{\sqrt{1-4t\tau}}\:dt\:d\tau\\
\label{normint}
\end{equation}
does not converge since by change of the variables
$$\rho = t\tau,\qquad d\tau = d\rho/t$$
one obtains
\begin{equation}
\int_0^{\infty}\frac{dt}{t}
\int_0^{\infty}\frac{1}{\sqrt{1-4\rho}}{(\rho)^{\gamma/4-1}}d\rho
\label{divnorm}
\,.
\end{equation}
Here the second integral converges for $0<\gamma<2$ while the first integral is divergent. The question is, is the norm of $|v>_{\gamma}$ finite and the norm of $|0>_k$ vanishing or the norm of $|v>_{\gamma}$ divergent and norm of $|0>_k$ finite? The point of view of this paper is that matrix elements built over $|v>_{\gamma}$ exist, while the standard vacuum $|0>$ having $\gamma=0$ is suspect.

\end{document}